\documentclass[11pt,a4paper]{article}

\usepackage[a4paper,margin=1in]{geometry}
\usepackage{setspace}
\usepackage[T1]{fontenc}
\usepackage[utf8]{inputenc}
\usepackage{lmodern}
\usepackage{CJK}

\usepackage{amsmath,amssymb,amsfonts}
\usepackage{bm}
\usepackage{mathtools}

\usepackage{graphicx}
\usepackage{subcaption}
\usepackage{booktabs}
\usepackage{multirow}
\usepackage{array}
\usepackage{tabularx}
\usepackage{siunitx}
\usepackage[dvipsnames]{xcolor}
\usepackage[colorlinks=true,
            linkcolor=MidnightBlue,
            citecolor=MidnightBlue,
            urlcolor=MidnightBlue]{hyperref}
\usepackage[nameinlink,capitalise]{cleveref}

\usepackage[numbers,sort&compress]{natbib}
\begin{document}
\begin{CJK}{UTF8}{gbsn}

\title{A Minimal Theoretical Model for Morphodynamic Selection in Soft Matter Systems}

\author{
Weihua Mu (牟维华) $^{1,2,*}$, Da Wang (王达)$^{3,4,5}$\\[0.5em]
\small $^{1}$ Zhejiang Key Laboratory of Soft Matter Biomedical Materials, Wenzhou Institute, \\ \small University of Chinese Academy of Sciences, Wenzhou, Zhejiang 325000, China\\
\small $^{2}$ Wenzhou Key Laboratory of Biomaterials and Engineering, Wenzhou Institute, \\ \small University of Chinese Academy of Sciences, Wenzhou, Zhejiang 325000, China\\
\small $^{3}$ Shenzhen Institutes of Advanced Technology, Chinese Academy of Sciences, Shenzhen, China \\
\small $^{4}$ University of Macau, Macau, China\\
\small $^5$ Shenzhen Key Laboratory for Exascale Engineering and Scientific Computing, Shenzhen, China}
%%}
\date{\today}

%\begin{document}

\maketitle

\begin{abstract}
Soft matter systems often display shape transitions, mode competition, and threshold-like dynamical responses driven by the coupling between mechanics, geometry, and biochemical or active fields. Here we present a minimal theoretical framework for describing such morphodynamic selection using a small number of physically interpretable order parameters. The model combines elastic relaxation, active driving, and nonlinear feedback, yielding distinct regimes corresponding to weak deformation, single-mode selection, and coupled shape transformation. We derive the governing equations, identify the relevant dimensionless control parameters, and show how the phase behavior can be organized in terms of threshold competition and effective coupling strengths. The framework provides a compact route for connecting analytical theory, numerical simulations, and experimentally observable shape dynamics in soft and biological matter.
\end{abstract}


%\vspace{1em}
%\noindent\textbf{Preprint note:} This manuscript is a preprint prepared for arXiv. A journal-formatted version may be prepared separately for submission to \textit{Soft Matter}.

% ============================================================
\section{Introduction}

Sequential switching between competing deformation modes is a common feature of deformable active systems. In such systems, a single internal or external drive can activate distinct shape responses at different thresholds, leading to ordered transitions, intermediate states, and competition-controlled switching behavior. Minimal models of this type are useful in soft matter because they reveal how complex morphodynamic phenomena can emerge from a small number of coupled amplitudes and effective control parameters. Living cells provide an important class of active soft objects in which such behavior can be studied, since their shape is determined by the interplay of cortical activity, internal mechanics, and geometrical constraints.

Animal-cell division is accompanied by large and highly organized changes of cell shape. Mitotic entry is associated with global cell rounding driven by increased cortical tension and hydrostatic pressure, whereas cytokinesis involves the formation and ingression of a localized equatorial furrow \cite{Taubenberger2020,FischerFriedrich2014}. RhoA-dependent cortical remodeling and its upstream regulators are centrally implicated in both mitotic cell-shape change and later cytokinetic contractility \cite{Maddox2003,Matthews2012,Chircop2014}. These observations motivate reduced mechano-chemical descriptions in which one regulatory pathway activates distinct mechanical shape programs in sequence. From a soft-matter perspective, this sequence may be viewed as ordered activation of two coupled deformation channels in an active, deformable object.

Recent biological and experimental work further supports the physical picture underlying such a reduced framework. First, mitotic entry is accompanied by increased cortical tension, intracellular pressure, and stiffening, which naturally favor a rounding-dominated morphology once the regulatory activity exceeds a threshold \cite{Taubenberger2020,FischerFriedrich2014,Gupta2022}. Second, recent studies of cytokinesis show that RhoA-centered cortical patterning and actin-network organization are required to maintain myosin II and RhoA at the cytokinetic furrow, thereby motivating a distinct furrowing response driven by the same activity but with delayed onset and nonlinear saturation \cite{Shah2024,Werner2024,Bement2024}. Third, recent work on cell and tissue mechanics emphasizes that mechanically regulated morphogenesis often proceeds through state transitions involving a small number of dominant modes, which supports the use of a reduced nonlinear description \cite{Mao2024}. Together, these observations suggest that the essential physics of the problem may be captured without resolving the full biochemical and structural complexity of the dividing cell.

However, the cell is not mechanically empty inside. The nucleus is the largest and stiffest organelle in many animal cells and acts as a major internal mechanical constraint that both resists deformation and transmits forces through the cytoskeleton and the nuclear envelope \cite{Lele2018,Kalukula2022,Miroshnikova2022}. More broadly, nuclear size, stiffness, and envelope mechanics shape whole-cell deformation, mechanotransduction, and the ability of cells to adapt to geometric confinement \cite{Lele2018,Kalukula2022,Lomakin2020,Alraies2024}. In soft-matter terms, the nucleus can therefore be regarded as unresolved internal structure that renormalizes the effective mechanical response of the whole cell. This suggests that a theory of cell-shape switching during division should not only include an activity variable promoting cortical contractility, but should also recognize that internal mechanical structure modulates the effective thresholds for global rounding and localized furrowing.

The present work deliberately narrows the modeling scope. We consider one mother cell only during the restricted interval from late interphase or mitotic entry through mitotic rounding and into early-to-late cytokinetic furrow ingression. The model does not address the full cell cycle, spindle mechanics, chromosome segregation, abscission, or the post-separation dynamics of daughter cells. Its final state should therefore be interpreted as a strongly constricted late-cytokinetic or pre-separation morphology rather than two independent daughter cells. Within this experimentally relevant window, the central theoretical question is not whether the system develops autonomous oscillations, but whether it undergoes threshold-driven switching between weakly deformed, rounded, and furrowed morphologies.

Motivated by this viewpoint, we formulate a minimal model that captures the interplay among a global rounding mode, a localized furrow mode, and a RhoA-like control activity. In the present reduced description, the effect of the nucleus is not introduced as an additional explicit dynamical variable. Instead, it is coarse-grained into the effective threshold and competition parameters, most notably the furrow threshold ratio \(p_f^\ast\) and the nonlinear mode-coupling coefficients \(\beta\) and \(\delta\). This treatment addresses a gap in the current literature on cell membrane deformation. Previous work has often regarded the nucleus either as a passive passenger or as a downstream biochemical target. By contrast, our framework treats unresolved internal structure, represented most directly by the nucleus, as an internal modulator of membrane-shape switching. The resulting theory supports a bistable and mode-competition picture in which one threshold controls mitotic rounding, a second threshold controls furrow initiation, and nonlinear coupling determines whether the transition is gradual or sharply switch-like.

In this way, the dividing-cell problem becomes a broader soft-matter problem of threshold-controlled morphodynamic selection in a deformable active medium. The model identifies a minimal active-shape architecture in which one internal activity sequentially activates two competing deformation programs, generates a robust intermediate state, and allows the influence of hidden internal mechanics to be incorporated through effective shifts of thresholds and couplings.

\section{Theory and Methods}

\begin{figure*}[htbp!]
	\centering
	%	%\begin{subfigure}{0.45\textwidth}
	%	%	\centering
	\includegraphics[width=0.9\textwidth]{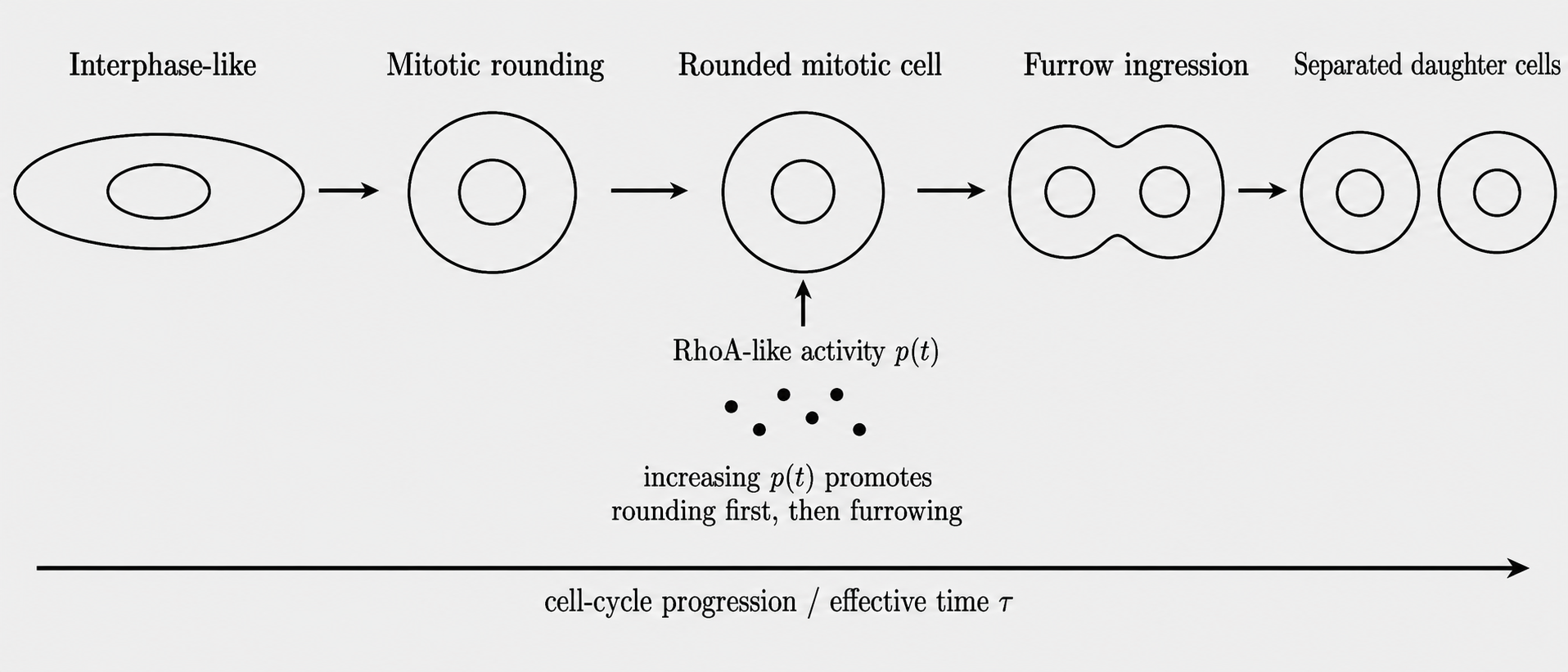} 
	\caption{Sketch illustration of the simplified mechano-chemical theory for cell-shape evolution during division.
		The cell progresses from an interphase-like morphology to mitotic rounding, then to a rounded mitotic state, followed by furrow ingression and daughter-cell formation.
		The protein activity \(p(\tau)\), interpreted as a RhoA-like regulator, drives the sequential activation of two shape modes: the rounding mode \(r(\tau)\) and the furrow mode \(f(\tau)\).
		The model predicts two characteristic thresholds, \(p=1\) for rounding onset and \(p=p_f^\ast\) for furrow initiation, thereby capturing the experimentally observed sequence of rounding before cytokinetic furrowing.}
	\label{fig:cycle}
\end{figure*}

\subsection{Theoretical Framework}

We introduce three dimensionless variables, $r(\tau)$, $f(\tau)$, and $p(\tau)$. The $r(\tau)$ denotes a global rounding mode, $f(\tau)$ denotes a furrow mode associated with equatorial constriction, and $p(\tau)$ denotes a normalized RhoA-like biochemical activity. More explicitly, $r(\tau)$ is a dimensionless measure of how close the mother cell is to a rounded mitotic morphology and may be extracted from cell height/width ratio, circularity, or sphericity; $f(\tau)$ is a dimensionless measure of furrow ingression and may be extracted from the neck radius or furrow depth; and $p(\tau)$ is a dimensionless activity variable for a RhoA-like contractility-promoting pathway and may be estimated from normalized fluorescence or activity readouts. Thus, the three variables represent, respectively, global cell rounding, equatorial constriction, and the intracellular biochemical drive that couples signaling to mechanics.

The governing equations are taken as
%~\cite{Maddox2003,Taubenberger2020,Gupta2022,Shah2024,Werner2024,Bement2024,Matthews2012,Chircop2014}
\begin{align}
	\dot r &= \mu_r(p-1)r-r^3-\beta r f^2,
	\label{eq:r}
	\\
	\dot f &= \mu_f(p-p_f^\ast)f-f^3-\delta r^2f+\xi p,
	\label{eq:f}
	\\
	\dot p &= \sigma s(\tau)-p+\rho_r r-\rho_f f,
	\label{eq:p}
\end{align}
where the overdot denotes differentiation with respect to dimensionless time $\tau$. The first equation describes the amplification and saturation of a rounding mode under RhoA-dependent cortical activation \cite{Maddox2003,Taubenberger2020,Gupta2022}. The second equation represents the delayed activation and nonlinear saturation of furrow formation, consistent with recent work on cytokinetic-furrow organization and mechanochemical feedback \cite{Shah2024,Werner2024,Bement2024}. The third equation models a minimal RhoA-like biochemical activity with activation, decay, and feedback from cell shape, in the spirit of Ect2/RhoA-controlled cortical remodeling \cite{Matthews2012,Chircop2014,Bement2024}. 

The parameters $\mu_r$ and $\mu_f$ quantify the protein-controlled growth of the rounding and furrow modes. The threshold ratio $p_f^\ast$ compares the furrow threshold to the rounding threshold. The coefficients $\beta$ and $\delta$ encode nonlinear competition between the two shape programs, $\xi$ represents direct protein-driven promotion of furrowing, and $\rho_r$ and $\rho_f$ represent feedback from shape to chemistry. The function $s(\tau)$ is a prescribed smooth activation schedule describing entry into the mechanically active stage of division.

The natural initial condition is a weakly deformed pre-mitotic state, $r(0)\ll 1$, $f(0)\approx 0$, and $p(0)$ is small, so the model starts from one mother cell that is neither strongly rounded nor furrowed.

The dominant nonlinear feature of Eqs.~(\ref{eq:r})--(\ref{eq:p}) is threshold-driven switching. For small $r$ and negligible furrowing, Eq.~(\ref{eq:r}) reduces to
\begin{equation}
	\dot r\approx \mu_r(p-1)r,
\end{equation}
which gives the rounding threshold $p=1$. Likewise, for small $f$ the furrow equation reduces to
\begin{equation}
	\dot f\approx \mu_f(p-p_f^\ast)f+\xi p,
\end{equation}
so the natural furrow threshold is $p=p_f^\ast$. Therefore, when $p_f^\ast>1$, the theory predicts a rounded-only intermediate regime, $1<p<p_f^\ast$, in which the cell has rounded but has not yet developed a strong furrow.

\section{Numerical Results}

For a representative example we fix the parameters as listed in Table~\ref{tab:parameters}.
\begin{table}[htbp]
	\caption{Representative dimensionless parameter set, grouped by physical function.}
	\label{tab:parameters}
	\centering
	\begin{tabular}{@{}llc@{}}
		\toprule
		Category & Parameter & Value \\
		\midrule
		\multirow{2}{*}{Growth} 
		& $\mu_r$ (rounding amplification) & $3.0$ \\
		& $\mu_f$ (furrowing amplification) & $2.0$ \\
		\midrule
		Threshold 
		& $p_f^\ast$ (furrow delay) & $1.55$ \\
		\midrule
		\multirow{2}{*}{Antagonism} 
		& $\beta$ (nonlinear coupling) & $0.8$ \\
		& $\delta$ (mode competition) & $0.6$ \\
		\midrule
		Forcing 
		& $\xi$ (protein-to-furrow) & $0.35$ \\
		\midrule
		\multirow{2}{*}{Feedback} 
		& $\rho_r$ (shape-to-chemistry, rounding) & $0.35$ \\
		& $\rho_f$ (shape-to-chemistry, furrowing) & $0.20$ \\
		\midrule
		Driving 
		& $\sigma$ (net chemical strength) & $1.8$ \\
		\bottomrule
	\end{tabular}
\end{table}
with a smooth activation schedule
\begin{equation}
	s(\tau)=\frac{1}{1+\exp[-(\tau-\tau_c)/w]},
\end{equation}
The function $s(\tau)$ is chosen as a logistic switching function to represent gradual activation of the biochemical drive during the relevant stage of the cell cycle. The parameter $\tau_c=8$ denotes the midpoint of activation, at which $s=1/2$, while the width parameter $w=2$ controls how sharply the activation turns on.

Starting from a weakly deformed pre-mitotic state with negligible furrow and low regulatory activity, \(\bigl(r(0),f(0),p(0)\bigr)=(0.03,0,0.15)\), the numerical trajectory shows a natural sequential progression of the division-associated shape program. In particular, the RhoA-like activity \(p(\tau)\) increases first under the prescribed cell-cycle drive, subsequently triggering rounding once it exceeds the threshold \(p=1\), while furrow formation emerges only later when \(p(\tau)\) reaches the higher threshold \(p=p_f^\ast\). The resulting sequence is
\begin{equation*}
	\text{weakly deformed} \to \text{rounded} \to \text{rounded plus furrowed}.
\end{equation*}

\begin{figure}[t]
	\centering
	\includegraphics[width=0.92\linewidth]{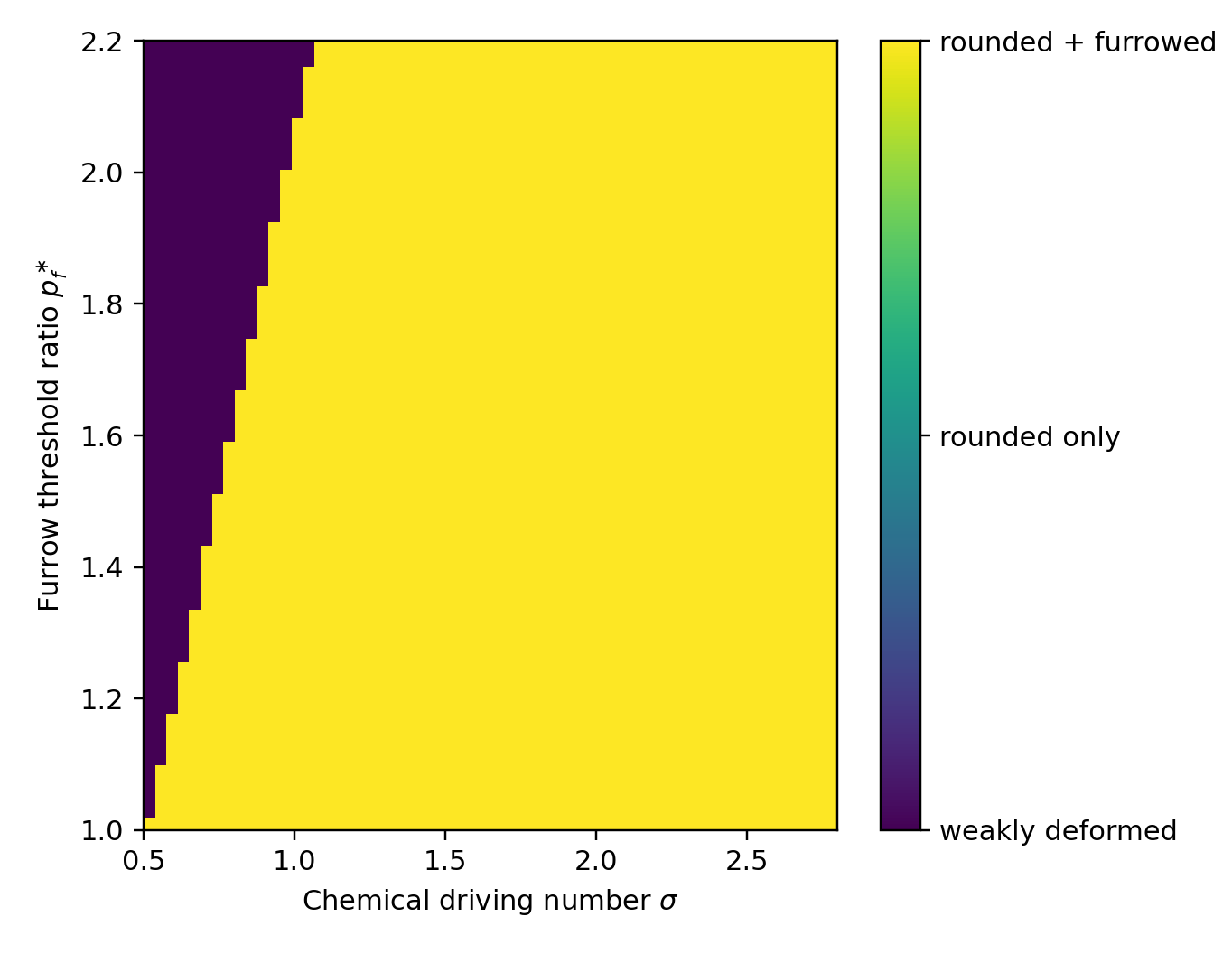}
	\caption{
		Phase map in the \((\sigma,p_f^\ast)\) plane for the reduced mechano-chemical model.
		Three classes of long-time behavior are identified: weakly deformed, rounded only, and rounded plus furrowed.
		Here \(\sigma\) is the chemical driving number and \(p_f^\ast\) is the threshold ratio for furrow activation.
		The diagram shows that increasing \(\sigma\) promotes shape evolution, whereas increasing \(p_f^\ast\) delays or suppresses furrow formation, thereby enlarging the rounded-only regime.
	}
	\label{fig:phase_sigma_pf}
\end{figure}

\begin{figure}[t]
	\centering
	\includegraphics[width=0.92\linewidth]{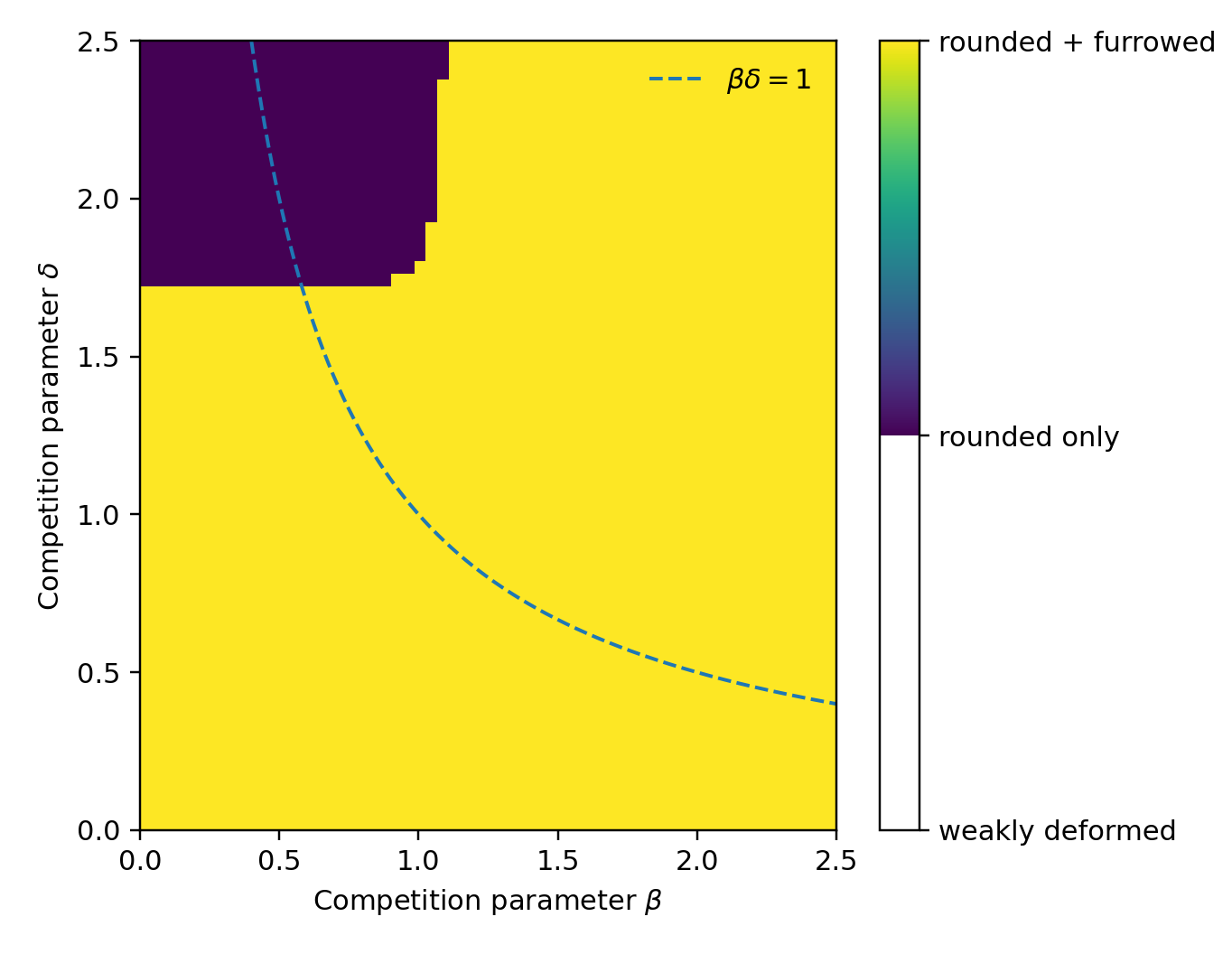}
	\caption{
		Phase map in the \((\beta,\delta)\) plane.
		The parameters \(\beta\) and \(\delta\) quantify the nonlinear competition between the rounding mode and the furrow mode.
		The dashed line \(\beta\delta=1\) is shown as a guide to the crossover between relatively weak competition and stronger antagonism.
		The diagram indicates that stronger competition sharpens the switching between weakly deformed, rounded-only, and rounded-plus-furrowed states.
	}
	\label{fig:phase_beta_delta}
\end{figure}
The zigzag-like boundaries between regions are mainly numerical rather than physical.
In the present calculations, the phase diagrams are obtained from a discrete scan in parameter space and each grid point is assigned to a region by threshold criteria on the long-time values of the shape variables, for example \(r_\infty\) and \(f_\infty\).
Near a true phase boundary, these long-time values vary only weakly and may lie close to the chosen classification thresholds, so a small change in the sampled parameter values can shift a point from one class to another.
As a result, the boundary appears jagged or staircase-like on a finite grid.
To discriminate regions more robustly on such non-smooth boundaries, one should either refine the parameter mesh locally, extend the integration time to ensure convergence to the asymptotic state, or test the sensitivity of the classification criteria, such as replacing the thresholds \(r_\infty=0.15\) and \(f_\infty=0.15\) by nearby values and confirming that the large-scale topology of the phase diagram remains unchanged.
Thus, the small zigzags should be interpreted as finite-resolution and thresholding effects, whereas the global organization of the phase map is the physically meaningful result. In the present work, the thresholds used to distinguish the three classes are chosen phenomenologically: a state is labeled weakly deformed if both \(r_\infty\) and \(f_\infty\) remain small, rounded only if \(r_\infty\) is appreciable but \(f_\infty\) remains small, and rounded plus furrowed if \(f_\infty\) also becomes appreciable.
These threshold values are not unique, but they are selected to separate the three morphologies in a transparent and reproducible way.

\begin{figure}[t]
	\centering
	\includegraphics[width=0.95\linewidth]{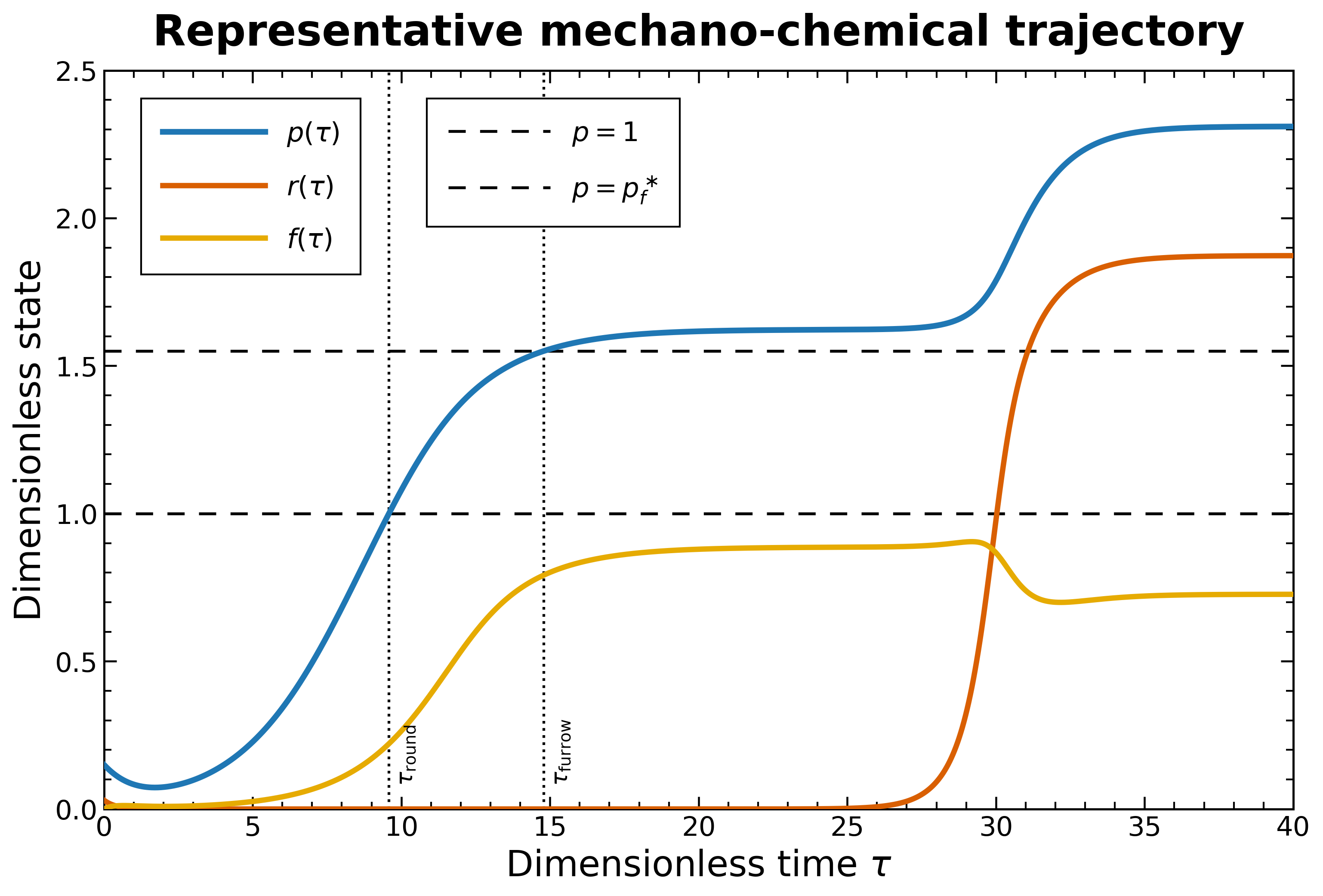}
	\caption{
		Representative mechano-chemical trajectory of the reduced bistable model.
		The dimensionless variables \(p(\tau)\), \(r(\tau)\), and \(f(\tau)\) denote, respectively, the RhoA-like biochemical activity, the rounding mode, and the furrow mode.
		The dashed horizontal lines indicate the thresholds \(p=1\) for rounding onset and \(p=p_f^\ast\) for furrow initiation, while the dotted vertical lines mark the corresponding threshold-crossing times.
	}
	\label{fig:timeseries_bistable_improved}
\end{figure}

\subsection{Sequential switching of shape programs and biological interpretation}

The main result of the present model is that the deformation of a single mother cell, within the restricted interval from mitotic entry to late cytokinesis, is better understood as a \emph{sequential switching process} than as an oscillatory one. This is seen most clearly in Fig.~\ref{fig:timeseries_bistable_improved}, where the biochemical activity \(p(\tau)\), the rounding mode \(r(\tau)\), and the furrow mode \(f(\tau)\) evolve in a definite temporal order. Starting from a weakly deformed pre-mitotic state, the prescribed cell-cycle drive first increases the RhoA-like activity \(p(\tau)\). When \(p(\tau)\) exceeds the first threshold, \(p=1\), the rounding mode grows rapidly, indicating the onset of global mitotic rounding. Only later, when \(p(\tau)\) reaches the higher threshold \(p=p_f^\ast\), does the furrow mode undergo strong amplification and the system move toward a rounded-plus-furrowed state.

This ordering is physically reasonable and biologically meaningful. In many animal cells, mitotic entry is accompanied by increased cortical tension, intracellular pressure, and stiffening, which collectively favor a rounded morphology \cite{FischerFriedrich2014,Taubenberger2020,Monster2021}. Cytokinetic furrow formation is a later event and requires more spatially organized contractility at the equator, with RhoA-centered cortical signaling playing a central role \cite{Matthews2012,Shah2024,Bement2024}. In the present theory, this temporal logic is captured by a minimal mechanism: a single biochemical variable \(p(\tau)\) acts through two different effective thresholds, one for rounding and one for furrowing. In this sense, the model does not simply reproduce two cell shapes; it explains why the two shape changes should occur in sequence.

Figure~\ref{fig:timeseries_bistable_improved} also shows that the furrow mode can deviate slightly from zero before \(p(\tau)\) crosses \(p_f^\ast\). This does not contradict the threshold picture. Rather, it reflects the direct forcing term in the furrow equation, which allows a weak precursor response before the onset of strong self-amplified furrow growth. Therefore, the threshold \(p=p_f^\ast\) should be interpreted as the onset of \emph{robust} furrow amplification, not necessarily the first instant at which a small equatorial indentation becomes measurable. This is again biologically plausible, since weak geometric asymmetries or small equatorial contractile signatures can appear before the furrow enters its strongly ingressive stage \cite{Chircop2014,Shah2024,Werner2024}.

Taken together, the time-series results support a bistable and mode-competition picture. The key question is not whether the cell oscillates between shapes, but when it leaves one shape program and commits to another. Within this restricted stage of division, the present model therefore provides a compact nonlinear explanation for the experimentally observed sequence in which RhoA-dependent cortical remodeling accompanies mitotic rounding and later supports cytokinetic furrow formation \cite{Maddox2003,Matthews2012,Monster2021}.

\subsection{Phase-diagram structure, mode competition, and quantitative interpretation}

The phase diagrams in Figs.~\ref{fig:phase_sigma_pf} and \ref{fig:phase_beta_delta} show how the model organizes cell-shape outcomes in parameter space. Figure~\ref{fig:phase_sigma_pf} displays three classes of long-time behavior in the \((\sigma,p_f^\ast)\) plane: weakly deformed, rounded only, and rounded plus furrowed. This figure is important because both control parameters have a direct mechanistic meaning. The parameter \(\sigma\) measures the effective strength of the biochemical drive, while \(p_f^\ast\) measures how much larger the biochemical activity must become for furrow formation than for rounding. Therefore, increasing \(\sigma\) corresponds to strengthening the intracellular program that promotes deformation, whereas increasing \(p_f^\ast\) makes furrow initiation more difficult relative to global rounding.

The large-scale organization of Fig.~\ref{fig:phase_sigma_pf} has a clear biological interpretation. When the chemical drive is weak, the system remains close to its weakly deformed state. At intermediate drive, the cell can round up without yet developing a strong furrow. At stronger drive, the rounded-plus-furrowed state becomes preferred. The existence of a substantial rounded-only region is one of the main conceptual results of this work. It shows that a rounded intermediate state is not merely an accidental transient, but a robust dynamical regime supported by the nonlinear structure of the model. This is exactly the behavior one would expect if mitotic rounding and cytokinetic furrowing are both controlled by the same regulatory pathway but require different effective levels of activation \cite{Maddox2003,Taubenberger2020,Gupta2022}.

Figure~\ref{fig:phase_beta_delta} addresses another question: how strongly do the two mechanical shape programs compete? Here \(\beta\) measures how strongly furrowing suppresses rounding, while \(\delta\) measures how strongly rounding suppresses furrowing. The dashed guide line \(\beta\delta=1\) is useful because it marks a simple crossover between relatively weak competition and stronger antagonism. For smaller values of the product \(\beta\delta\), the two modes can coexist more easily, and the transition from rounding to furrowing is smoother. For larger values, the two shape programs oppose one another more strongly, and the switch between them becomes sharper and more abrupt.

This competition-based interpretation is important biologically. Shape change during division does not result from independent forces acting in isolation. Rather, global cortical contractility favors rounding, whereas localized equatorial constriction favors furrowing, and these tendencies can either coexist or compete. The present model suggests that the balance between them can itself act as a control principle. When competition is weak, mixed morphologies are easier to sustain; when competition is strong, the cell tends to remain longer in one dominant shape state before switching more sharply to the next. In this way, Fig.~\ref{fig:phase_beta_delta} gives a nonlinear-science interpretation of how mechanical antagonism can regulate the sharpness of morphological transitions during division \cite{Bement2024,Mao2024,Werner2024}.

A further point worth emphasizing is that the fine zigzag-like details of some phase boundaries should not be overinterpreted. These small irregularities mainly reflect the finite resolution of the numerical parameter scan and the practical thresholding used to classify long-time states. Near a true boundary, the asymptotic values of \(r(\tau)\) and \(f(\tau)\) can vary only weakly, so a small change in the sampled grid or in the classification criterion can move a point from one class to another. The physically meaningful result is therefore the overall topology of the phase diagrams, not the exact pixel-level shape of the boundaries. The important message is the robust separation of the three morphological regimes and the clear role of the control parameters in organizing them.

\subsection{Nuclear modulation, significance of the reduced framework, and limitations}

An important conceptual point of the present work is that it highlights the role of the nucleus even though the model does not contain an explicit nuclear state variable. In many dividing cells, the nucleus is the largest and often the stiffest intracellular structure. Its size, stiffness, and envelope integrity influence how easily cortical forces can be transmitted into global cell deformation. In the present reduced framework, these nuclear effects are absorbed into effective thresholds and couplings rather than represented by a separate dynamical subsystem. This is a physically meaningful coarse-graining: a stiffer or larger nucleus should resist deformation transfer more strongly, prolong the rounded-but-not-yet-furrowed state, and raise the effective barrier to deep constriction. By contrast, a more compliant nucleus should facilitate the transmission of cortical forcing into membrane-shape change \cite{Lele2018,Kalukula2022,Miroshnikova2022,Lomakin2020}.

The requested recent references strengthen this interpretation from several complementary directions. Venturini \emph{et al.} showed that the nucleus can function as a mechanical sensor for cell-shape change, supporting the broader idea that nuclear mechanics can feed back on whole-cell morphology \cite{Venturini2020}. Monster \emph{et al.} demonstrated that junctional mechanotransduction helps coordinate mitotic rounding with tissue integrity, which supports the view that rounding is not only biochemically driven but also mechanically constrained \cite{Monster2021}. Niethammer reviewed the components and mechanisms of nuclear mechanotransduction, reinforcing the idea that nuclear properties can influence how intracellular forces are interpreted and redistributed \cite{Niethammer2021}. Dantas \emph{et al.} further showed that nuclear tension can directly regulate mitotic entry, which makes it natural to interpret the nucleus as a factor that can shift the effective thresholds in a reduced dynamical model \cite{Dantas2022}. Finally, Lammerding \emph{et al.} emphasized that nuclear mechanics should be treated as an active component of cell mechanobiology rather than as a passive geometric inclusion \cite{Lammerding2022}. Taken together, these works support the central interpretation of the present theory: the nucleus can modulate the \emph{threshold structure} of membrane deformation even if it is represented only implicitly.

This viewpoint helps define the novelty of the present study. Much of the existing literature on cell-shape change during division emphasizes either biochemical signaling or cortex-based mechanics. The present model adds a third layer of interpretation. Even without introducing a full nuclear mechanics model, it shows how the nucleus can be understood as an internal modulator of the threshold structure of membrane deformation. In other words, nuclear mechanics does not merely change the amount of deformation; it can also reshape the nonlinear decision structure that determines whether the cell remains weakly deformed, becomes rounded, or proceeds into furrowing. This provides a useful conceptual bridge between cell membrane deformation, cortical regulation, and nuclear mechanobiology \cite{Kalukula2022,Niethammer2021,Lammerding2022,Alraies2024}.

The limitations of the theory should also be stated clearly. First, the model does \emph{not} describe the full cell cycle. Its intended scope is limited to a single mother cell during the interval from late interphase or mitotic entry to late cytokinesis. Second, the final state should be interpreted as a deeply constricted pre-separation morphology, not as two independently evolving daughter cells. A true post-separation theory would require either two daughter-cell state vectors or a geometric framework with topology change. Third, because nuclear effects are represented only implicitly, the theory cannot distinguish among specific microscopic mechanisms such as lamina softening, nuclear-envelope rupture, chromatin reorganization, or nucleus--cortex decoupling. Finally, the model is phenomenological: its parameters are effective quantities chosen to capture the dominant nonlinear structure of the deformation process, rather than a one-to-one mapping to specific molecular rate constants.

These limitations, however, are closely related to the strength of the approach. By reducing the problem to one biochemical activity and two competing shape modes, the model makes the nonlinear logic of the transition unusually transparent. The present study therefore does not claim to be a full mechanistic description of cell division. Instead, it provides a compact and physically readable framework for understanding how biochemical regulation, effective nuclear modulation, and mechanical competition can organize the sequential switching of membrane-shape programs during division.

\section{Discussion}

\subsection{Sequential switching of deformation modes}

The central result of the present model is that the shape evolution of a single mother cell, within the restricted interval from mitotic entry to late cytokinesis, is more naturally interpreted as a \emph{sequential switching process} than as an oscillatory one. This is seen most directly in Fig.~\ref{fig:timeseries_bistable_improved}, where the control activity \(p(\tau)\), the global rounding mode \(r(\tau)\), and the localized furrow mode \(f(\tau)\) evolve in a definite temporal order. Starting from a weakly deformed pre-mitotic state, the prescribed cell-cycle drive first increases the RhoA-like activity \(p(\tau)\). When \(p(\tau)\) exceeds the first threshold, \(p=1\), the rounding mode grows rapidly, indicating the onset of global mitotic rounding. Only later, when \(p(\tau)\) reaches the higher threshold \(p=p_f^\ast\), does the furrow mode undergo strong amplification and the system move toward a rounded-plus-furrowed morphology.

This ordering is physically reasonable and biologically meaningful. In many animal cells, mitotic entry is accompanied by increased cortical tension, intracellular pressure, and stiffening, which collectively favor a rounded morphology \cite{FischerFriedrich2014,Taubenberger2020,Monster2021}. Cytokinetic furrow formation is a later event and requires more spatially organized contractility at the equator, with RhoA-centered cortical signaling playing a central role \cite{Matthews2012,Shah2024,Bement2024}. In the present theory, this temporal logic is captured by a minimal active-shape architecture: a single internal activity \(p(\tau)\) acts through two distinct effective thresholds, one for a global deformation channel and one for a localized constriction channel. In this sense, the model does not merely reproduce two cell shapes; it explains why the two shape changes should occur in sequence.

Figure~\ref{fig:timeseries_bistable_improved} also shows that the furrow mode can deviate slightly from zero before \(p(\tau)\) crosses \(p_f^\ast\). This does not contradict the threshold picture. Rather, it reflects the direct forcing term in the furrow equation, which permits a weak precursor response before the onset of strong self-amplified furrow growth. The threshold \(p=p_f^\ast\) should therefore be interpreted as the onset of \emph{robust} furrow amplification, not necessarily the first instant at which a small equatorial indentation becomes measurable. This interpretation is again biologically plausible, since weak geometric asymmetries or incipient equatorial contractile signatures can appear before the furrow enters its strongly ingressive stage \cite{Chircop2014,Shah2024,Werner2024}.

From a soft-matter viewpoint, this behavior is naturally interpreted as sequential activation of two coupled deformation modes in a deformable active object. The global mode is activated first because it is the lower-threshold response of the system, whereas the localized constriction mode requires both stronger activity and a favorable balance of nonlinear coupling. The main question is therefore not whether the cell oscillates between shapes, but when it leaves one deformation program and commits to another. Within this restricted stage of division, the present model provides a compact nonlinear explanation for the experimentally observed sequence in which RhoA-dependent cortical remodeling accompanies mitotic rounding and later supports cytokinetic furrow formation \cite{Maddox2003,Matthews2012,Monster2021}.

\subsection{Phase-diagram structure and morphodynamic selection}

The phase diagrams in Figs.~\ref{fig:phase_sigma_pf} and \ref{fig:phase_beta_delta} show how the model organizes cell-shape outcomes in parameter space. Figure~\ref{fig:phase_sigma_pf} displays three classes of long-time behavior in the \((\sigma,p_f^\ast)\) plane: weakly deformed, rounded only, and rounded plus furrowed. Both control parameters have direct physical meaning. The parameter \(\sigma\) measures the effective strength of the internal drive, while \(p_f^\ast\) measures how much larger the activity must become for furrow formation than for rounding. Therefore, increasing \(\sigma\) corresponds to strengthening the internal program that promotes deformation, whereas increasing \(p_f^\ast\) makes localized constriction more difficult relative to global rounding.

The large-scale organization of Fig.~\ref{fig:phase_sigma_pf} has a clear morphodynamic interpretation. When the drive is weak, the system remains close to its weakly deformed state. At intermediate drive, the cell can round up without yet developing a strong furrow. At stronger drive, the rounded-plus-furrowed state becomes preferred. The existence of a substantial rounded-only region is one of the main conceptual results of the work. It shows that the intermediate morphology is not simply an accidental transient, but a robust dynamical regime supported by the nonlinear structure of the model. This is exactly the kind of behavior expected if global rounding and localized furrowing are both controlled by the same internal activity but require different effective levels of activation \cite{Maddox2003,Taubenberger2020,Gupta2022}.

This point is important in the context of soft matter because intermediate morphologies need not correspond to incompletely evolved states. Instead, they may represent selected configurations stabilized over finite parameter ranges by threshold separation and nonlinear coupling. In the present system, the rounded-only state is such an example: it is a morphologically distinct regime that persists because one deformation channel has been activated while the second remains below its effective onset.

Figure~\ref{fig:phase_beta_delta} addresses a different question: how strongly do the two shape programs compete? Here \(\beta\) measures how strongly furrowing suppresses rounding, while \(\delta\) measures how strongly rounding suppresses furrowing. The dashed guide line \(\beta\delta=1\) serves as a simple indicator of the crossover from relatively weak competition to stronger antagonism. For smaller values of the product \(\beta\delta\), the two modes can coexist more easily, and the transition from rounding to furrowing is smoother. For larger values, the two deformation channels oppose one another more strongly, and the switch between them becomes sharper and more abrupt.

This competition-based interpretation is physically useful because shape change in an active deformable object does not result from independent responses acting in isolation. Rather, different deformation tendencies coexist, interact, and constrain one another. In the biological realization studied here, global cortical contractility favors rounding, whereas localized equatorial constriction favors furrowing. In more general soft-matter language, the system contains a global deformation channel and a localized constriction channel, and their nonlinear interaction determines whether the shape evolution is gradual or switch-like. The present model therefore suggests that the balance between competing modes can itself act as a control principle for morphodynamic selection.

A further point worth emphasizing is that the small zigzag-like details of some phase boundaries should not be overinterpreted. These fine-scale irregularities mainly reflect the finite resolution of the numerical parameter scan and the practical thresholding used to classify long-time states. Near a true boundary, the asymptotic values of \(r(\tau)\) and \(f(\tau)\) can vary only weakly, so a small change in the sampled grid or in the classification criterion can move a point from one class to another. The physically meaningful result is therefore the overall topology of the phase diagrams, not the exact pixel-level boundary shape. What matters is the robust separation of weakly deformed, rounded-only, and rounded-plus-furrowed regimes, together with the clear role of the control parameters in organizing them.

\subsection{Internal mechanical structure and the role of the nucleus}

An important conceptual point of the present work is that it highlights the role of the nucleus even though the model does not contain an explicit nuclear state variable. In many dividing cells, the nucleus is the largest and often the stiffest intracellular structure. Its size, stiffness, and envelope integrity influence how easily cortical forces can be transmitted into global cell deformation. In the present reduced framework, these nuclear effects are absorbed into effective thresholds and couplings rather than represented by a separate dynamical subsystem. This is a physically meaningful coarse-graining: a stiffer or larger nucleus should resist deformation transfer more strongly, prolong the rounded-but-not-yet-furrowed state, and raise the effective barrier to deep constriction. By contrast, a more compliant nucleus should facilitate the transmission of cortical forcing into membrane-shape change \cite{Lele2018,Kalukula2022,Miroshnikova2022,Lomakin2020}.

Several recent studies reinforce this interpretation from complementary directions. Venturini \emph{et al.} showed that the nucleus can function as a mechanical sensor for cell-shape change, supporting the broader idea that nuclear mechanics can feed back on whole-cell morphology \cite{Venturini2020}. Monster \emph{et al.} demonstrated that junctional mechanotransduction helps coordinate mitotic rounding with tissue integrity, which supports the view that rounding is not only biochemically driven but also mechanically constrained \cite{Monster2021}. Niethammer reviewed the components and mechanisms of nuclear mechanotransduction, reinforcing the idea that nuclear properties can influence how intracellular forces are interpreted and redistributed \cite{Niethammer2021}. Dantas \emph{et al.} further showed that nuclear tension can directly regulate mitotic entry, which makes it natural to interpret the nucleus as a factor that can shift effective thresholds in a reduced dynamical model \cite{Dantas2022}. Lammerding \emph{et al.} emphasized that nuclear mechanics should be treated as an active component of cell mechanobiology rather than as a passive geometric inclusion \cite{Lammerding2022}. Taken together, these works support the central interpretation of the present theory: unresolved internal structure, represented most directly by the nucleus, can modulate the \emph{threshold structure} of membrane deformation even if it is represented only implicitly.

This viewpoint helps define the novelty of the present study. Much of the existing literature on cell-shape change during division emphasizes either biochemical signaling or cortex-based mechanics. The present model adds a third layer of interpretation. Even without introducing a full nuclear mechanics model, it shows how the nucleus can be understood as an internal modulator of the threshold structure of membrane deformation. In other words, internal mechanics does not merely change the amount of deformation; it can also reshape the nonlinear decision structure that determines whether the cell remains weakly deformed, becomes rounded, or proceeds into furrowing. This creates a useful conceptual bridge between membrane deformation, cortical regulation, and hidden internal mechanical constraints \cite{Kalukula2022,Niethammer2021,Lammerding2022,Alraies2024}.

In soft-matter terms, the nucleus may therefore be regarded as unresolved internal architecture that renormalizes the effective response of the whole active object. This interpretation is appealing because it allows internal mechanical structure to be incorporated at the reduced level through threshold and coupling shifts, without the need to specify a detailed constitutive theory for every intracellular component. The price of this simplification is that different microscopic mechanisms are combined into a small number of effective parameters. The benefit is that the resulting model remains transparent and easily interpretable.

\subsection{Scope, limitations, and broader significance}

The limitations of the theory should be stated clearly. First, the model does \emph{not} describe the full cell cycle. Its intended scope is limited to a single mother cell during the interval from late interphase or mitotic entry to late cytokinesis. Second, the final state should be interpreted as a deeply constricted pre-separation morphology, not as two independently evolving daughter cells. A true post-separation theory would require either two daughter-cell state vectors or a geometric framework with topology change. Third, because nuclear effects are represented only implicitly, the theory cannot distinguish among specific microscopic mechanisms such as lamina softening, nuclear-envelope rupture, chromatin reorganization, or nucleus--cortex decoupling. Finally, the parameters are phenomenological: they are effective quantities chosen to capture the dominant nonlinear structure of the deformation process, rather than one-to-one molecular rate constants.

These limitations, however, are closely related to the strength of the approach. By reducing the problem to one internal activity and two competing deformation modes, the model makes the logic of sequential shape selection unusually transparent. The work is therefore not intended as a full mechanistic description of cell division. Instead, it provides a compact and physically readable framework in which threshold separation, nonlinear competition, and hidden internal mechanics together determine the morphodynamic response of an active deformable system.

More broadly, the present model shows how a small nonlinear system can capture a large amount of organized shape behavior. One control activity activates a first deformation program, then a second one; the two programs compete; and internal structure shifts the effective switching conditions. This combination of ordered thresholds, mode competition, and unresolved internal mechanics is likely relevant beyond the specific biological realization studied here. At the same time, the dividing-cell problem provides an experimentally grounded setting in which these soft-matter ideas can be interpreted, visualized, and tested.

\section{Conclusions}

In this work, we developed a minimal nonlinear model for the shape evolution of a single mother cell during the restricted interval from mitotic entry to cytokinetic furrow ingression. The model is written in terms of three dimensionless variables: a global rounding mode \(r(\tau)\), a localized furrow mode \(f(\tau)\), and a RhoA-like control activity \(p(\tau)\). Despite its simplicity, the model captures a clear and physically meaningful sequence of shape transitions.

The first main result is the identification of two effective thresholds. The threshold \(p=1\) marks the onset of mitotic rounding, whereas the threshold \(p=p_f^\ast\) marks the onset of strong furrow amplification. When \(p_f^\ast>1\), the theory naturally produces a rounded-only intermediate regime before furrowing begins. This provides a simple and transparent explanation for why one internal activity can drive two different deformation programs in sequence.

The second main result is that the system is best understood as a bistable selector of shape modes, not as an autonomous oscillator. The numerical results show transitions among weakly deformed, rounded, and rounded-plus-furrowed morphologies. The phase diagram in the \((\sigma,p_f^\ast)\) plane identifies when each of these outcomes occurs, while the phase diagram in the \((\beta,\delta)\) plane shows how nonlinear competition controls whether the switch between shape programs is smooth or sharp. In soft-matter terms, the model therefore describes threshold-controlled morphodynamic selection in a deformable active object.

The third main result is the internal-mechanics interpretation of the reduced theory. Although the nucleus is not introduced as an explicit dynamical variable, the model shows how nuclear size, stiffness, and envelope integrity can be understood as shifting the effective thresholds and couplings that govern membrane deformation. In this sense, the present work goes beyond a purely protein-centered description and highlights \emph{implicit nuclear control of bistable membrane-shape switching} as a useful conceptual perspective for active cell morphodynamics.

The scope of the model is deliberately limited. It does not describe the full cell cycle, spindle mechanics, chromosome segregation, abscission, or the emergence of two fully separated daughter cells. Its final state should instead be interpreted as a strongly constricted late-cytokinetic or pre-separation morphology of a single mother cell. Within this clearly defined scope, however, the present theory provides a compact, transparent, and testable framework for understanding how internal activity, nonlinear competition, and unresolved internal mechanics can jointly organize successive membrane-shape transitions.

Overall, the novelty of the present work lies in combining a reduced mechano-chemical description, bistable nonlinear analysis, and an internal-mechanics interpretation into one unified picture. This makes the model easy to interpret physically, straightforward to compare qualitatively with experiments, and useful as a starting point for future extensions that include explicit nuclear dynamics, topology change, or more detailed active constitutive descriptions.

% ============================================================
\section*{Author contributions}

%{\cblue 待补充}
%
%\section*{Author contributions}
W.M. contributed to the conception of the study, model formulation, numerical analysis, and writing of the manuscript. D.W. contributed to the numerical analysis and writing of the manuscript. All authors approved the final version.

\section*{Data availability}
No experimental datasets were generated or analysed during the current study. Numerical codes used to generate the results and figures are available from the corresponding author upon reasonable request.

\end{CJK}
%\section*{Acknowledgements}

% ============================================================
% Bibliography
% Put references in references.bib and compile with BibTeX.
\bibliography{soft_matter_revised_refs}

% ============================================================
%\appendix
%
%\section{Additional derivation}
%\label{app:derivation}
%
%This appendix may contain derivations of the reduced equations, nondimensionalization steps, or stability analysis.
%
%\section{Numerical details}
%\label{app:numerical_details}
%
%This appendix may contain solver settings, parameter tables, convergence tests, or additional simulations.

\end{document}